\documentclass[letterpaper,10 pt,twocolumn,conference,numbers,sortcompress]{IEEEtran}
\IEEEoverridecommandlockouts

\usepackage{cite}
\usepackage{amsmath,amssymb,amsfonts}
\usepackage{algorithmic}
\usepackage{graphicx}
\usepackage{textcomp}
\usepackage{xcolor}
\usepackage{url}
\usepackage{hyperref}
\usepackage{blindtext}
\usepackage{caption}
\usepackage{booktabs}
\usepackage{multirow}
\usepackage[section]{placeins}
\usepackage{caption}
\hypersetup{
    colorlinks = true,
    linkcolor = blue,
    urlcolor = blue,
}
\def\BibTeX{{\rm B\kern-.05em{\sc i\kern-.025em b}\kern-.08em
    T\kern-.1667em\lower.7ex\hbox{E}\kern-.125emX}}
\begin{document}

\title{\LARGE \bf{Origin-Destination Pattern Effects on Large-Scale Mixed Traffic Control via Multi-Agent Reinforcement Learning}}

\author{\IEEEauthorblockN{1\textsuperscript{st} Muyang Fan}
\IEEEauthorblockA{\textit{Department of Computer Science } \\
\textit{University of Memphis }\\
Memphis, USA\\
mfan1@memphis.edu}
\\
\IEEEauthorblockN{3\textsuperscript{rd} Shuai Li}
\IEEEauthorblockA{\textit{Department of Civil \& Coastal Engineering} \\
\textit{University of Florida}\\
Gainesville, USA \\
shuai.li@essie.ufl.edu}
\and
\IEEEauthorblockN{2\textsuperscript{nd} Songyang Liu}
\IEEEauthorblockA{\textit{Department of Civil \& Coastal Engineering} \\
\textit{University of Florida}\\
Gainesville, USA\\
liusongyang@ufl.edu}
\\
\IEEEauthorblockN{4\textsuperscript{th} Weizi Li}
\IEEEauthorblockA{\textit{Min H. Kao Department of Electrical Engineering and Computer Science} \\
\textit{University of Tennessee, Knoxville}\\
Knoxville, USA \\
weizili@utk.edu}

}

\maketitle

\begin{abstract}
Traffic congestion remains a major challenge for modern urban transportation, diminishing both efficiency and quality of life. While autonomous driving technologies and reinforcement learning (RL) have shown promise for improving traffic control, most prior work has focused on small-scale networks or isolated intersections. Large-scale mixed traffic control, involving both human-driven and robotic vehicles, remains underexplored.
In this study, we propose a decentralized multi-agent reinforcement learning framework for managing large-scale mixed traffic networks, where intersections are controlled either by traditional traffic signals or by robotic vehicles. We evaluate our approach on a real-world network of 14 intersections in Colorado Springs, Colorado, USA, using average vehicle waiting time as the primary measure of traffic efficiency. We are exploring a problem that has not been sufficiently addressed: Is large-scale Multi-Agent Traffic Control (MTC) still feasible when facing time-varying Origin-Destination (OD) patterns?
\end{abstract}

\section{Introduction}
Traffic congestion is a major challenge for modern urban transportation, reducing traffic efficiency and quality of life. Prolonged vehicle idling not only increases energy consumption and economic losses but also intensifies environmental pollution, especially in cities with high vehicle emission concentrations during certain times. Traffic congestion can result in substantial economic losses. In port cities like Chittagong, severe congestion can cost up to \$2.01 million per day~\cite{congestion}. Furthermore, a study by INRIX and the Centre for Economics and Business Research projected that traffic-related delays could cost the UK, France, Germany, and the US a combined \$300 billion by 2030~\cite{BHARADWAJ20173538}. In addition, in congested conditions, vehicle efficiency drops sharply, leading to a substantial increase in emissions per mile. Prolonged exposure to these emissions, intensified by the effects of urban heat islands, poses growing risks to public health, including an increased incidence of asthma, cardiovascular diseases, and increased psychological stress and anxiety~\cite{trafficdisease,pollution}.


To ease traffic congestion, researchers have turned to autonomous driving technologies. With advancements accelerating, more vehicles now feature autonomous capabilities and are undergoing real-world trials. However, a complete transition to fully autonomous transport is unlikely to happen overnight. Instead, human-driven vehicles (HVs) and robot vehicles (RVs) will share the roads for a long period, creating a mixed traffic environment that demands careful coordination.


Traffic control has advanced considerably in recent years, with reinforcement learning (RL) emerging as a powerful tool in complex traffic scenarios. RL-based vehicle control strategies often outperform traditional traffic signal methods and have recently been extended to large-scale mixed traffic environments, especially for urban intersection management \cite{RLt1,RLt2}. Among these, multi-agent reinforcement learning (MARL) stands out for its strong performance and novel approach, attracting growing research interest \cite{MARL1,MARL2}.

We conduct experiments on a large-scale traffic network of 14 intersections using the Simulation of Urban MObility (SUMO)\cite{lopez2018microscopic}. In this network, some intersections are managed by traffic signals, while others are controlled by RVs equipped with RL capabilities. Our performance evaluation metric is average vehicle waiting time, which measures traffic efficiency by averaging the total time vehicles remain motionless near an intersection. We conduct experiments under various OD flow configurations and RV penetration rates. The results indicate that variations in OD configurations, when combined with appropriate RV penetration rates, can lead to significant differences in congestion levels. These findings suggest that adjusting the primary OD flow directions could serve as a viable strategy for alleviating urban traffic congestion. Such an approach opens up new avenues for dynamic traffic management and offers valuable insights for the future development of intelligent transportation systems. Results demonstrate that strategically adjusting major OD flow patterns can effectively influence congestion, offering a new pathway for enhancing urban mobility. This indicates that large-scale Multi-Agent Traffic Control (MTC) remains highly feasible even under time-varying OD patterns. Our work fills a gap in the current research and provides a more practical perspective for the future implementation of MTC in modern transportation systems in the real world. The code of our work is available at \url{https://github.com/cgchrfchscyrh/MixedTrafficControl_IROS}.

\section{Related Work}
Managing urban intersections has grown more challenging with the coexistence of autonomous and human-driven vehicles. Traditional methods like fixed-time and actuated signals have been widely used under stable conditions~\cite{NACTO, Greenlight}, but often fail to handle the dynamic, complex nature of modern traffic~\cite{gholamhosseinian2022comprehensive}. Optimization-based methods, such as integer programming and rule-based strategies, improve efficiency but face scalability limits in large networks~\cite{qadri2020state}.
In contrast, learning-based methods adapt in real time to changing traffic conditions. Following this direction, we propose an RL-based vehicle control strategy to enhance traffic flow.

Previous studies on unsignalized intersection control have focused primarily on connected and autonomous vehicles (CAVs). Early work introduced a multi-agent system in which CAVs reserve space-time slots on a first-come, first-served (FCFS) basis~\cite{dresner2008multiagent}. Later research expanded FCFS to CAV platoons~\cite{jin2013platoon} and proposed dynamic gap adjustments to reduce conflicts~\cite{chen2022improved}. Decentralized approaches, like consensus-based trajectory planning~\cite{mirheli2019consensus} and energy optimization~\cite{malikopoulos2018decentralized}, have also been explored, but most assume a fully autonomous environment. In contrast, we address mixed traffic by assuming only RVs can communicate across intersections, offering a more realistic and scalable control strategy for real-world urban networks.

Recent studies have shown that RL not only improves the influence of RVs on HVs~\cite{RLt1,RLt2}, but also enhances efficiency in multi-objective lane-change decision-making~\cite{RLt3}. Human-in-the-loop RL approaches have further demonstrated strong safety performance~\cite{RLt4}, contributing to reduced congestion and improved overall traffic efficiency. MARL has emerged as a particularly innovative and effective strategy for alleviating congestion at the network level~\cite{MARL1,MARL2,Marl6}. Its ability to adapt to non-stationary traffic environments~\cite{MARL3}, proactively coordinate connected autonomous vehicles in complex scenarios~\cite{Marl4}, and reduce collision risks~\cite{Marl5} has been well documented. However, while MARL-based traffic control has shown great promise, previous research has largely overlooked how variations in traffic flow OD patterns influence congestion severity. Unlike prior studies that focus primarily on vehicle-level or intersection-level control under static traffic conditions, our work extends MARL to large-scale mixed traffic networks by systematically varying RV penetration rates and OD distributions. 

\section{{Methodology}}
\subsection{Mixed Traffic Control at Intersections Based on RL}
To tackle the challenge of coordinating mixed traffic at unsignalized intersections, we frame the problem as a \textit{Partially Observable Markov Decision Process (POMDP)}, represented by the tuple:
\[
\mathcal{M} = \left( \mathcal{S}, \mathcal{A}, \mathcal{T}, \mathcal{R}, \mathcal{O}, \mathcal{Z}, \mu_0, \gamma \right)
\]

Where:
\begin{itemize}
    \item $\mathcal{S}$: State space (set of all possible traffic states),
    \item $\mathcal{A}$: Action space (e.g., \textit{Go} or \textit{Stop}),
    \item $\mathcal{T}(s'|s,a)$: Transition probability function,
    \item $\mathcal{R}(s,a)$: Reward function,
    \item $\mathcal{O}$: Observation space (for robot vehicles),
    \item $\mathcal{Z}(o|s)$: Observation probability,
    \item $\mu_0$: Initial state distribution,
    \item $\gamma \in [0, 1]$: Discount factor.
\end{itemize}

In this framework, each RV operates at discrete time steps and selects an action $a_t \in A$ based on its current state using a shared policy $\pi_\theta(a_t \mid s_t)$. Once an action is taken, the environment evolves to a new state $s_{t+1}$, and the vehicle receives an immediate reward $r_t$. Although all RVs act independently, they follow the same learned policy, aiming to maximize the cumulative discounted return:

\begin{equation}
    G_t = \sum_{j=t}^{T} \gamma^{j-t} r_j.
\end{equation}

Our action space is binary.  Each RV faces a binary decision at every time step: \textbf{Stop} or \textbf{Go}. A \textbf{Stop} action holds the vehicle at the intersection entrance, while \textbf{Go} allows it to proceed. The RV makes this decision based on an observation vector $o_t$, which encodes traffic information across all directions:
\begin{equation}
o_t = \bigoplus_{d \in D} \langle q_d, \tau_d \rangle \bigoplus_{d \in D} \sigma_d,
\end{equation}
where $D$ is the set of all incoming directions, $q_d$ is the number of queued vehicles from direction $d$, $\tau_d$ is their average waiting time, and $\sigma_d$ indicates whether direction $d$ currently occupies the intersection.

To encourage smooth traffic flow and minimize conflicts, we design a reward function that combines local efficiency with penalties for unsafe behavior. The total reward at time $t$ is given by:
\begin{equation}
r_t = \beta \cdot r_t^{\text{local}} + r_t^{\text{penalty}}
\end{equation}

The local reward is defined based on the action space discussed above :
\[
r_t^{\text{local}} =
\begin{cases}
+\tau_d & \text{if action is } \text{Go} \\
-\tau_d & \text{if action is } \text{Stop}
\end{cases}
\]

The penalty term accounts for conflicts:
\[
r_t^{\text{penalty}} =
\begin{cases}
-1 & \text{if a conflict occurs} \\
0 & \text{otherwise}
\end{cases}
\]

\subsection{Vehicle Behavior Modeling}
HVs are modeled using the Intelligent Driver Model (IDM)~\cite{treiber2000congested}, which determines acceleration based on surrounding traffic. RVs follow a hybrid strategy: adhere to IDM when more than 30 meters from an intersection and switch to a learned control policy within the intersection zone.

Under this policy, an RV instructed to Go accelerates at its maximum allowable rate. If instructed to Stop, it decelerates based on its current speed $v$ and the remaining distance to the intersection $d_{\text{int}}$, using the formula $-v^2 / (2d_{\text{int}})$.

\subsection{Mixed Control of Intersections}
We implement two distinct intersection control strategies: one based on RL-coordinated robot vehicles (RVs) and the other on traffic signals. In RV-controlled intersections, traffic lights are deactivated, allowing RVs to manage traffic flow autonomously. In our traffic network, two intersections are managed by RVs and 12 by traffic signals. For comparison, the baseline configuration represents all 14 intersections that are managed by traffic signals. This design allows studying the impact of transferring traffic signal control to RVs on large-scale traffic.

\subsection{Evaluation Metric}
The primary metric in this study is average waiting time ($\overline{W}$), defined as the total time a vehicle remains stationary within controlled areas. This widely used indicator reflects traffic efficiency: lower values signify reduced congestion and smoother flow~\cite{zhang2020using, greguric2020application, wang2024learning}. We calculate this metric by averaging the waiting times of all vehicles in a specific direction or at a given intersection. The overall average waiting time, $\overline{W}$, is obtained by dividing the total waiting time of all vehicles by their count.

\section{{Experiments and Results}}
\subsection{Experiment set-up}\label{AA}
Our experimental setup consists of two unsignalized intersections managed by RVs and 12 signalized intersections controlled by traffic lights. For training, we apply the Rainbow DQN algorithm~\cite{hessel2018rainbow}, conducting 1,000 episodes per configuration. Rainbow DQN is a reinforcement learning algorithm that integrates several improvements over the basic Deep Q-Network (DQN). It extends the original DQN by incorporating advancements such as Double DQN, Dueling Network Architecture, and Distributional RL. These enhancements collectively improve the algorithm's ability to provide adaptive policies for heterogeneous agents, optimal decision-making, efficient learning under sparse rewards, and robustness in uncertain scenarios. ~\cite{hessel2018rainbow} In this work, we adopt a simplified yet efficient version of Rainbow DQN and apply it within a multi-agent setting, where all robotic vehicles (RVs) share a single policy network but make decisions independently. To implement Rainbow DQN, we use the standard DQN configuration with dueling and double Q-learning mechanisms enabled, and we conduct training within a multi-agent Ray framework. Additionally, we define the objective function for both the policy and the Q-function to support the learning process of Rainbow DQN as follows.

For the objective function defined below, the goal of each agent is to maximize the expected cumulative discounted reward over time:
\begin{equation}
\max_{\pi_\theta} \mathbb{E} \left[ \sum_{j=t}^{T} \gamma^{j-t} r_j \right]
\end{equation}

where:
\begin{itemize}
    \item $\pi_\theta(a_t | o_t)$ is the shared policy of all RVs, parameterized by $\theta$.
    \item $r_j$ is the reward received at time step $j$.
    \item $\gamma$ is the discount factor (set to $0.99$ in this paper).
    \item $T$ is the episode horizon (1000 seconds).
\end{itemize}
For the Q-function it is defined as:

\begin{equation}
Q^\pi(o_t, a_t) = \mathbb{E}_{\pi} \left[ R_t \mid o_t, a_t \right], \quad \text{where} \quad R_t = \sum_{k=0}^{T-t} \gamma^k r_{t+k}
\end{equation}

Here:
\begin{itemize}
    \item $o_t$ denotes the local observation received by the agent at time step $t$;
    \item $a_t$ is the action selected at time $t$, where $a_t \in \{\texttt{Go}, \texttt{Stop}\}$;
    \item $r_{t+k}$ is the reward obtained at future step $t+k$;
    \item $\gamma \in (0,1)$ is the discount factor applied to future rewards (set to $\gamma = 0.99$);
    \item $\pi$ represents the shared stochastic policy followed by all agents after time $t$.
\end{itemize}
Because the environment is modeled as a decentralized partially observable Markov decision process (Dec-POMDP), each agent estimates its Q-value based solely on partial local information, without access to the full environment state. To learn the Q-function, the Rainbow DQN algorithm is used, which incorporates several state-of-the-art DQN improvements:
\begin{itemize}
    \item Double Q-learning: mitigates overestimation bias by separating the action selection (from the online network) and value evaluation (using the target network).

    \item Dueling Network Architecture: decomposes the Q-value function into two separate estimators:
    \begin{equation}
    Q(o, a) = V(o) + \left( A(o, a) - \frac{1}{|\mathcal{A}|} \sum_{a' \in \mathcal{A}} A(o, a') \right)
    \end{equation}
    where $V(o)$ is the estimated state value, and $A(o, a)$ is the advantage of taking action $a$ under observation $o$.

    \item Distributional RL: Instead of estimating a scalar Q-value, it learns the full probability distribution of possible returns using 51 discrete support atoms.
\end{itemize}

RV penetration rates are varied at 25\%, 50\%, 75\%, and 100\%. Training times range from 8 to 16 hours per configuration on a machine equipped with an Intel Core™ i9-14900K CPU and an NVIDIA RTX 4080 GPU. We use a prioritized replay buffer with $\alpha = 0.5$ and a capacity of 50,000. The neural network architecture consists of three hidden layers, each with 512 neurons, and employs 51 atoms for distributional output. Key hyperparameters include a discount factor of 0.99, a mini-batch size of 32, a learning rate of 0.0005, and a control zone radius of 30 meters. Each trained policy is evaluated over 100 simulation runs, each lasting 1,000 seconds, with 8,000 vehicles running in the network. Final performance metrics are computed by averaging results across these evaluations. 

We define 12 OD pairs covering key traffic directions: South–North (SN), North–South (NS), North–East (NE), East–North (EN), North–West (NW), West–North (WN), South–East (SE), East–South (ES), South–West (SW), West–South (WS), West–East (WE), and East–West (EW). These OD pairs are combined across different configurations to support our experimental analysis.

The eight experimental configurations are as follows: \begin{enumerate}[]

\item \textbf{NS + SN 70\%}: 70\% of traffic flows between North–South and South–North; the remaining 30\% is distributed among other directions.

\item \textbf{NS + SN 90\%}: 90\% of traffic flows between North–South and South–North; 10\% is distributed among other directions.

\item \textbf{NW + WN 70\%}: 70\% of traffic flows between North–West and West–North; the remaining 30\% is distributed among other directions.

\item \textbf{NW + WN 90\%}: 90\% of traffic flows between North–West and West–North; 10\% is distributed among other directions.

\item \textbf{WE + EW 70\%}: 70\% of traffic flows between West–East and East–West; 30\% is distributed among other directions.

\item \textbf{Equally Distributed}: Traffic is evenly distributed across all twelve OD pairs, with each direction accounting for approximately 8.33\% of the total traffic.

\item \textbf{NE + SW 70\%}: 70\% of traffic flows from North–East and South–West; the remaining 30\% is distributed among other directions.

\item \textbf{SE + NW 70\%}: 70\% of traffic flows from South–East and North–West; the remaining 30\% is distributed among other directions.

\end{enumerate}

These eight experiments are designed to explore how different OD flow combinations affect the overall network’s average waiting time through the following six comparative analyses: \begin{enumerate}

\item[a] \textbf{Experiments 1 vs. 2}: Assess the impact of increasing the proportion of North–South and South–North traffic on network efficiency and mixed traffic performance.

\item[b] \textbf{Experiments 3 vs. 4}: Analyze how varying the dominance of North–West and West–North traffic influences traffic efficiency and mixed traffic dynamics.

\item[c] \textbf{Experiments 1 vs. 3}: Compare the effects of primary flows in straight (North–South) versus turning (North–West) directions on mixed traffic performance.

\item[d] \textbf{Experiments 5 vs. 1}: Examine the impact of different straight-line traffic flows (West–East vs. North–South) on mixed traffic efficiency.

\item[e] \textbf{Experiment 6 vs. others}: Evaluate whether evenly distributed traffic across all directions yields better network performance compared to concentrated flow patterns.

\item[f] \textbf{Experiments 7 vs. 8}: Investigate how different OD flow patterns affect the overall average waiting time.

\end{enumerate}

\begin{figure*}
    \centering
    \includegraphics[scale=0.223]{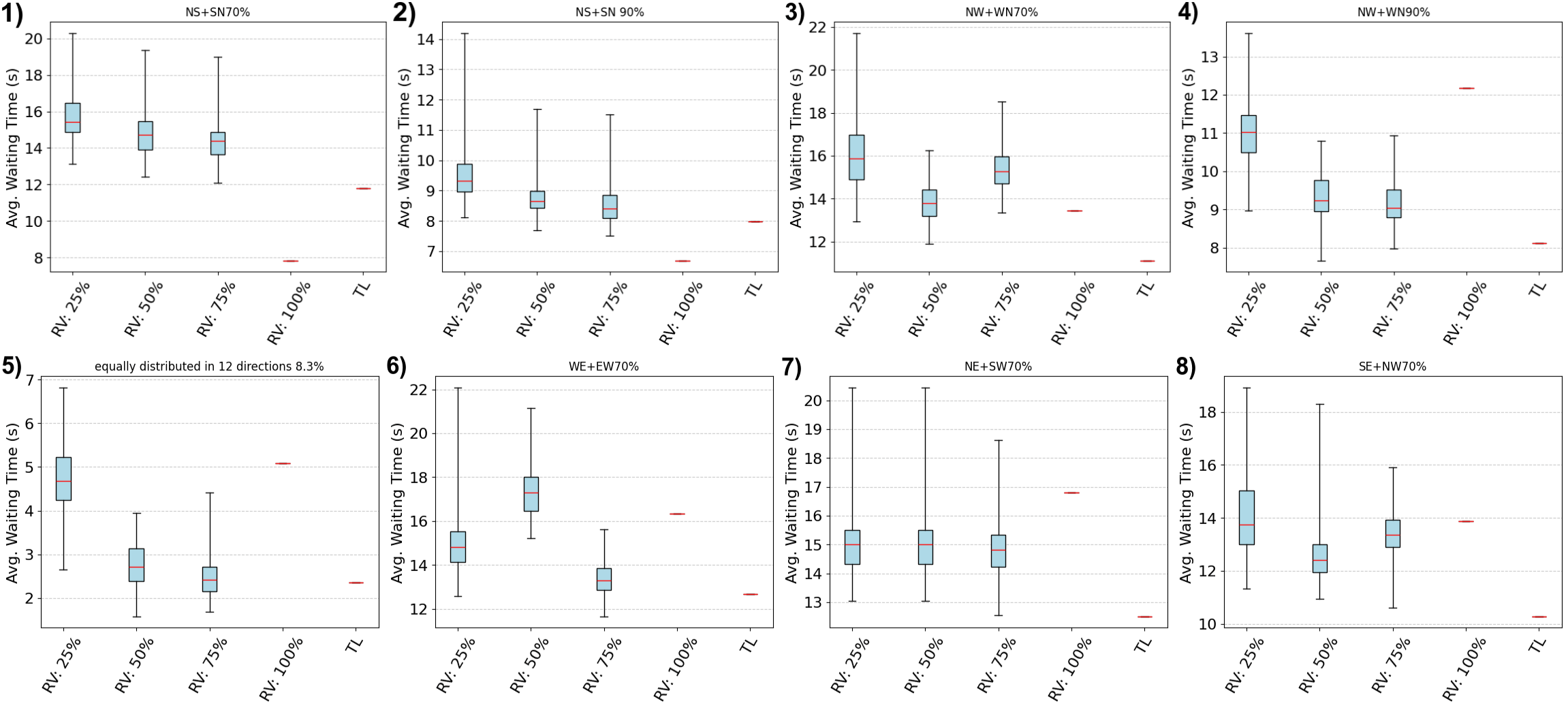}
    \vspace{-20pt}
    \caption{\small{Average waiting time~$\overline{W}~(s)$ of the eight experiments from 1) to 8). In 1), NS+SN is the main traffic flow direction, accounting for 70\% of the traffic. In 2), NS+SN remains the main traffic flow direction, but the proportion increases to 90\%, decreasing average waiting time. In 3), NW+WN is the main traffic flow direction, with 70\% of the traffic concentrated. Compared to Experiment 1, the difference in average waiting time is not significant. In 4), NW+WN remains the main traffic flow direction, with the traffic proportion increased to 90\%, resulting a decrease in average waiting time. 5) features traffic evenly distributed across 12 directions, which significantly reduces the average waiting time of the traffic network compared to the other experiments. In 6), WE+EW is the main traffic flow direction. Compared to 1), the average waiting time is slightly longer. 7) focuses on NE+SW as the main traffic flow direction. In 8), SE+NW is the main traffic flow direction, and compared to Experiment 7, the average waiting time is slightly lower. }}
    \label{fig1}     
\vspace{-1.5em}
\end{figure*}

\subsection{Results and Analysis}
The experimental results in Fig.~\ref{fig1} and Table~\ref{tab1} show that a uniform traffic distribution across all 12 OD pairs significantly reduces average waiting time, outperforming all other configurations.

\begin{table}[ht] 
\centering
\scalebox{1}{ 
\begin{tabular}{c|ccccc} 
\toprule
& \multicolumn{5}{c}{RV rate} \\
\cmidrule(l){2-6} 
 & 100\% & 75\% & 50\% & 25\% & Baseline \\
\cmidrule(l){2-6} 
{\centering Experiment} & \multicolumn{5}{c}{$\overline{W} (s)$} \\
\midrule
1. NS+SN 70\% & 14.69 & 16 & 15.09 & 7.83 & 11.8 \\
\midrule
2. NS+SN 90\% & 6.69 & 7.89 & 11.35 & 8.56 & 7.97 \\
\midrule
3. NW+WN 70\% & 14.6 & 17.8 & 12.48 & 13.46 & 11.12 \\
\midrule
4. NW+WN 90\% & 12.18 & 9.35 & 9.21 & 10.34 & 8.11 \\
\midrule
5. 12 Dir. Uniform. & 5.09 & \textbf{2.22} & 2.66 & 4.71 & 2.36 \\
\midrule
6. WE+EW 70\% & 16.35 & 15.43 & 16.50 & 14.96 & 12.66 \\
\midrule
7. NE+SW 70\% & 16.81 & 15.62 & 14.17 & 18.11 & 12.51 \\
\midrule
8. SE+NW 70\% & 13.87 & 13.96 & 12.38 & 13.31 & 10.28 \\
\bottomrule
\end{tabular}}
\caption{\small{Average waiting time~$\overline{W}$ of the eight experiments from 1) to 8). ``12 Dir Uniform'' means 12 directions in uniform distribution. The experimental results indicate that a uniform distribution of traffic across all 12 OD directions significantly reduces average waiting times within the traffic network compared to other configurations. Specifically, when the primary traffic flow follows straight-line OD patterns such as NS+SN, increasing the emphasis on the direction from 70\% to 90\% leads to a decrease in average waiting time. Similarly, for a non-straight-line OD pattern like NW+WN, a higher traffic concentration (from 70\% to 90\%) in this direction also reduces waiting time. When comparing non-straight-line OD patterns (e.g., NW+WN) to straight-line patterns (e.g., NS+SN), there is no major difference. Among straight-line OD configurations, the NS+SN pattern contributes better performance than the WE+EW pattern overall. When evaluating different combinations of OD patterns, the SE+NW pattern demonstrates better outcomes than the SW+NE pattern.}}
\label{tab1}
\end{table}

Detailed findings from the six comparative analyses are as follows. \begin{enumerate} \item[a)] For straight-line OD flows (NS+SN), increasing the concentration from 70\% to 90\% leads to an overall reduction in average waiting time. When the RV penetration rate is 25\%, the average waiting time for the NS+SN 90\% is 0.73~$s$ higher than that of the NS+SN 70\%. When RV penetration rates are 50\%, 75\%, and 100\%, it shows the decrease in average waiting time by 3.74~$s$, 8.11~$s$, and 8~$s$, respectively. \item [b)] A similar trend is observed for the non-straight-line turning flows (NW+WN), where a 90\% concentration also improves traffic efficiency compared to 70\%. At RV penetration rates of 0.25, 0.5, 0.75, and 1.0, the configuration emphasizing NW+WN at 90\% consistently resulted in lower average waiting time than the NW+WN 70\% configuration. Specifically, the reductions in average waiting time were 3.12~$s$, 3.27~$s$, 8.45~$s$, and 2.42~$s$, respectively. \item[c)] Comparing NS+SN 70\% with NW+WN 70\%, neither consistently outperforms the other. NS+SN yields lower waiting times at RV rates of 25\% and 75\%, the average waiting time is reduced by 5.63~$s$ and 1.8~$s$, respectively. While NW+WN performs better at 50\% and 100\%, the average waiting time is reduced by 2.61~$s$ and 0.09~$s$, respectively. Consequently, this does not indicate that prioritizing straight-line traffic direction (NS+SN) yields better performance over non-straight pattern (NW+WN). \item[d)] Among straight-line OD patterns, NS+SN 70\%, the vertical direction outperforms WE+EW, the horizontal direction, in reducing average waiting time. At RV penetration rates of 25\%, 50\%, and 100\%, the average waiting time for the NS+SN 70\% configuration was reduced by 7.13~$s$, 1.41~$s$, and 1.66~$s$, respectively, compared to the WE+EW 70\%. 
\item[e)] The configuration with evenly distributed traffic (8.3\% per direction) achieves the best overall performance, with the lowest average waiting time of 2.22~$s$ at an RV penetration rate of 75\%. Specifically, at RV penetration rates of 25\%, 75\%, and 100\%, the evenly distributed 12-direction configuration has lower waiting times by 3.85~$s$, 5.67~$s$, and 1.6~$s$, respectively, than NS+SN 90\%, the second lowest. 
\item[f)] Between two configurations comprised of entirely different patterns, SE+NW results in lower waiting times than NE+SW, showing a clear advantage. At RV penetration rates of 25\%, 50\%, 75\%, and 100\%, the SE+NW 70\% configuration consistently has a decreased waiting time than NE+SW 70\% configuration, with reductions of 4.9~$s$, 1.79~$s$, 1.66~$s$, and 2.94~$s$, respectively.\end{enumerate}

These results highlight the importance of both traffic flow distribution and directionality in managing congestion. Notably, adjusting RV penetration alone does not guarantee consistent improvements across all scenarios. While higher RV rates generally help, the optimal performance is achieved when traffic is evenly distributed, especially at 75\% RV penetration- average waiting time of 2.22~$s$. In contrast, the NE + SW configuration in 25\% RV penetration yields the worst result, with an average waiting time of 18.11~$s$. A uniform OD flow distribution is most effective in minimizing congestion.

\section{Conclusion and Future Work}
We apply MARL to large-scale mixed traffic intersection control, conducting eight quasi-realistic experiments on a real-world urban network in Colorado Springs, USA, comprising 14 intersections. By exploring different OD flow patterns, we evaluate their influence on network-wide traffic performance. Results show that a uniform distribution of traffic across all 12 OD directions consistently yields the lowest average waiting time. Additionally, several OD configurations with directional emphasis also lead to improved performance, highlighting the potential of OD pattern design as an effective strategy for mitigating urban traffic congestion and proving that large-scale Multi-Agent Traffic Control (MTC) still holds potential when facing time-varying OD patterns.

This work has several limitations. We only consider a fixed configuration of two unsignalized and twelve signalized intersections, without evaluating alternative distributions (e.g., 6 unsignalized + 8 signalized). We also focus exclusively on standard intersection geometries, excluding more complex layouts such as roundabouts and highway ramps. Furthermore, the influence of RV penetration rates shows non-linear patterns that merit deeper analysis. Future work will address these limitations by incorporating a wider range of intersection types and geometries, and by systematically analyzing the effects of varying RV penetration rates. These extensions aim to improve the robustness and generalizability of MARL-based traffic control strategies for real-world urban environments. The transition from intersections fully controlled by traffic signals to a system without signals is a gradual process. Our research can serve as a theoretical foundation for pilot testing of signal-free intersections, starting with a single intersection that operates without traffic lights.


\section*{Acknowledgment}

This research is funded by the National Science Foundation (NSF) via Grant 2129003, 2231710, and 2153426. The authors gratefully acknowledge NSF’s support. The authors would also like to thank NVIDIA and the Tickle College of Engineering at University of Tennessee, Knoxville for their support.

\bibliographystyle{IEEEtran}
\bibliography{ref}


\end{document}